\documentclass[twocolumn,twoside,slac_two]{revtex4}
\usepackage{graphicx}
\usepackage{fancyhdr}
\begin{document}

\title{Review of $\theta_{13}$ and Prospects for Double Chooz}

%

\author{I. Gil-Botella}
\affiliation{Centro de Investigaciones Energ\'eticas, Medioambientales y Tecnol\'ogicas (CIEMAT), Madrid, Spain}

\begin{abstract}
The Double Chooz reactor neutrino experiment will be the next detector to search for a non vanishing $\theta_{13}$ mixing angle with unprecedented sensitivity, which might open the way to unveiling CP violation in the leptonic sector. The measurement of this angle will be based on a precise comparison of the antineutrino spectrum at two identical detectors located at different distances from the Chooz nuclear reactor cores in France. Double Chooz is particularly attractive because of its capability to measure $sin^2(2\theta_{13})$ to 3$\sigma$ significance if $sin^2(2\theta_{13})$ $>$ 0.05 or to exclude $sin^2(2\theta_{13})$ down to 0.03 at 90\% C.L. for $\Delta$m$^2$ = 2.5 $\times$ 10$^{-3}$ eV$^2$ in three years of data taking with both detectors. The installation of the far detector started in May 2008 and the first neutrino results are expected in 2009. The advantages of reactor neutrino experiments to measure the $\theta_{13}$ mixing angle are described in this article and in particular, the design, current status and expected performance of the Double Chooz detector.

\end{abstract}

\maketitle

\thispagestyle{fancy}


\section{Physics Motivations}

The neutrino oscillation phenomenon has been clearly established by the study of solar, atmospheric, reactor and beam neutrinos. The PMNS mixing matrix relates the three neutrino mass eigenstates to the three neutrino flavor eigenstates. This can be parametrized by three mixing angles ($\theta_{12}$, $\theta_{13}$, $\theta_{23}$) and one CP violating phase $\delta_{CP}$ (if neutrinos are Dirac particles). During the last years, tremendous progress has been achieved in the experimental field trying to measure the values of $\theta_{ij}$ and the two squared mass differences $\Delta m^{2}_{ij} = m^{2}_{i} - m^{2}_{j}$ which govern the oscillation probabilities. The two mass differences and the $\theta_{12}$ and $\theta_{23}$ mixing angles have been measured with good precision \cite{GonzalezGarcia:2007ib}. However, the $\theta_{13}$ angle, the sign of $\Delta m^{2}_{31}$ (mass hierarchy) and the $\delta_{CP}$ phase are still unknown.

In particular, only an upper limit on the value of $\theta_{13}$ has been established indicating that the angle is very small compared to the other mixing angles. A three-flavor global analysis of the existing data provides a constraint on $\theta_{13}$ being $sin^2\theta_{13}$ $<$ 0.028 at 90\% C.L \cite{Schwetz:2006dh}. This limit is essentially dominated by the result obtained by the CHOOZ reactor experiment \cite{Apollonio:2002gd} in France. This experiment measured the fraction of $\bar{\nu_e}$'s surviving at a distance of 1.05 km from the reactor cores to be R = 1.01 $\pm$ 2.8\% (stat) $\pm$ 2.7\% (syst). This result was mainly limited by the systematic uncertainties induced by the imperfect knowledge of the neutrino production and interaction.

The measurement of this angle is of fundamental interest not only for the final understanding of neutrino oscillations but because it determines the possibilities to observe CP violation in the leptonic sector with the forthcoming neutrino experiments.

\section{Measuring $\theta_{13}$ at Nuclear Reactors}

The information on the $\theta_{13}$ mixing angle can be essentially obtained from accelerator or nuclear reactor neutrino experiments. The long baseline accelerator experiments measure the appearance of $\nu_e$'s in a $\nu_{\mu}$ beam generated at long distance from the detector. The $\nu_{\mu}$ $\to$ $\nu_e$ transition depends on several oscillation parameters like the CP phase and the sign of $\Delta m^{2}_{31}$. Moreover, they can also be sensitive to matter effects due to the long
 baselines. Therefore, the measurement of $\theta_{13}$ will be affected by correlations and degeneracies between parameters and the sensitivity of the accelerator experiments to this parameter will be reduced. 

On the other hand, reactor experiments are unique to provide an unambiguous determination of $\theta_{13}$. Nuclear reactors are very intense sources of $\bar{\nu_e}$'s coming from the $\beta$-decay of the neutron-rich fission fragments of fuel elements (U and Pu).

Reactor neutrino experiments will look for the disappearance of $\bar{\nu_e}$'s with energies extending up to 10 MeV over distances of the order of kilometers (short baselines) to maximize the disappearance probability. Due to the low energy of the emitted electron antineutrinos, only disappearance measurements can be performed. The survival probability of $\bar{\nu_e}$'s emitted by a nuclear power station can be written as:

\begin{equation}
\begin{array}{ll}
P(\bar{\nu_e} \to \bar{\nu_e}) & = 1 - sin^2(2\theta_{13})sin^2\frac{\Delta m^{2}_{31}L}{4E} \\
\\
&  - cos^4\theta_{13}sin^2(2\theta_{12})sin^2\frac{\Delta m^{2}_{21}L}{4E} \\
\\
&  + 2 sin^2\theta_{13}cos^2\theta_{13}sin^2\theta_{12} \\
\\
& \left( cos\frac{(\Delta m^{2}_{31}-\Delta m^{2}_{21})L}{2E} - cos\frac{\Delta m^{2}_{31}L}{2E} \right)
\end{array}
\label{osc-prob}
\end{equation}

\noindent where $E$ is the neutrino energy and $L$ the distance from the source to the detector. Only the first term is relevant for short baselines and therefore the oscillation amplitude is proportional to $sin^2\theta_{13}$ and independent of the CP phase and the mass hierarchy. 

A clean measurement of $\theta_{13}$ can be performed with reactor neutrino experiments since they do not suffer, unlike accelerator experiments, from degeneracies and correlations between different oscillation parameters. Since they are short baseline experiments, they are not affected by matter effects.

Reactor antineutrinos are detected through the inverse beta decay $\bar{\nu_e}$ + p $\to$ n + $e^+$. The energy threshold for this reaction is 1.806 MeV. The signature of the neutrino interaction is the coincidence of the photons from the prompt signal due to the $e^+$ annihilation and the delayed signal ($\Delta$t $\sim$ 30 $\mu$s) from the neutron capture. Many liquid scintillator $\bar\nu$ experiments use scintillator loaded with Gadolinium in their fiducial volume because of its large neutron capture cross section and high total $\gamma$ yield of 7-8 MeV.

The neutrino spectrum results from the convolution of the neutrino flux and the cross section. The spectrum peaks at $\sim$ 4 MeV. Reactor experiments will look for a suppression of the $\bar{\nu_e}$ rate and a distortion of the spectrum shape due to oscillations. 

The signature of a neutrino interaction can be mimicked by two kind of background events: accidentals and correlated. The accidental background corresponds to the coincidence of a positron-like signal coming from natural radioactivity of the surrounding environment or of the detector materials, with a neutron induced by cosmic muon spallation in the surrounding rock and captured in the detector. The correlated background are events that mimic both parts of the coincidence signal. They come from fast neutrons produced by cosmic muons, which can produce proton-recoils in the target scintillator (misidentified as e$^+$) and then are captured after thermalisation or from the $\beta$ - n decay of long-lived cosmogenic radioisotopes ($^9$Li, $^8$He) produced by muon interactions in the scintillator.

Some of the largest systematic uncertainties of the CHOOZ reactor experiment are related to the accuracy to which the original neutrino flux and spectrum are known. In order to improve the CHOOZ sensitivity to the $\theta_{13}$ mixing angle, a relative comparison between two or more identical detectors located at different distances from the power plant is required. The first one, which is located at few hundred meters from the nuclear cores, monitors the neutrino flux and spectrum before neutrinos oscillate. The second detector located 1-2 km away from the cores searches for a departure from the overall 1/L$^2$ behavior of the neutrino energy spectrum. At the same time, the statistical error can also be reduced by increasing the exposure and the fiducial volume of the detector. Moreover, backgrounds can be further reduced with a better detector design, using veto detectors and external shields against muons and external radioactivity.

Several nuclear reactor neutrino experiments are foreseen to measure the $\theta_{13}$ mixing angle: Double Chooz \cite{Ardellier:2006mn} in France, Daya Bay \cite{DayaBay} in China, RENO \cite{reno} in South Korea and Angra \cite{angra} in Brazil. In the next section the first of these experiments is explained in detail.

\section{The Double Chooz Experiment}

The Double Chooz experiment \cite{Ardellier:2006mn} will improve our knowledge on the $\theta_{13}$ mixing angle within a competitive time scale and for a modest cost. The Double Chooz collaboration is composed by institutes of Brazil, France, Germany, Japan, Russia, Spain, UK and USA. 

\begin{figure}[h]
\centering
\includegraphics[width=85mm]{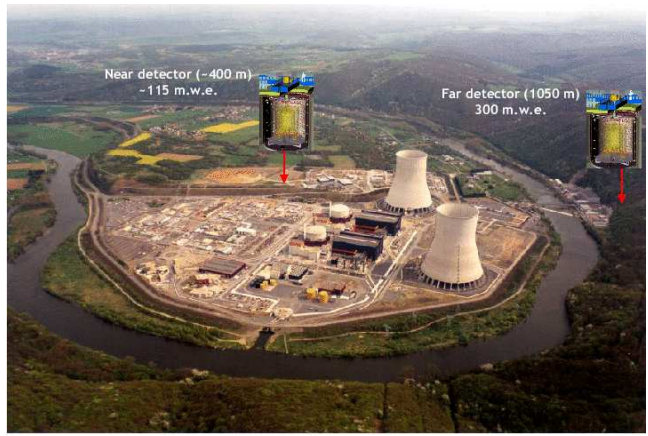}
\caption{\label{chooz}Location of the two Double Chooz detectors in the Chooz-B nuclear power plant.}
\end{figure}

The experiment is being installed in the Chooz-B nuclear power plant in the Northeast of France. The maximum operating thermal power of the two cores is 8.54 GW$_{th}$. The far detector is located at 1050 meters from the cores in the same laboratory used by the CHOOZ experiment (see Fig.~\ref{chooz}). It provides a quickly prepared and well-shielded (300 meters of water equivalent (m.w.e.)) site with near-maximal oscillation effect. A second identical detector (near detector) will be installed at 400 m away from the reactor cores to cancel the lack of knowledge of the neutrino spectrum and reduce the systematic errors related to the detector. For the near site, a tunnel will be excavated under a small natural hill (overburden 115 m.w.e.) and a near lab will be equipped.

In addition to the neutrino oscillation studies, Double Chooz aims to study the possibility of using this kind of detector to verify the non-proliferation of nuclear weapons in the framework of the International Atomic Energy Agency missions. A near detector could be able to provide a precise measurement of the reactor power and isotopic composition by looking at the energy spectrum and rate of the emitted antineutrinos, following its evolution with time.

\subsection{Experimental design}

\begin{figure}[h]
\centering
\includegraphics[width=85mm]{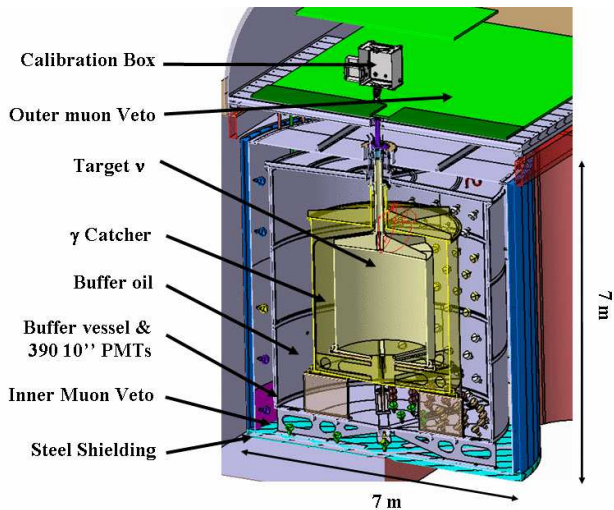}
\caption{\label{detector}The Double Chooz Detector Design.}
\end{figure}

The CHOOZ detector design can also be optimized in order to reduce backgrounds. The Double Chooz detectors (see Fig.~\ref{detector}) consist of concentric cylinders and an outer plastic scintillator muon veto. The innermost volume (``target'') contains about 10 tons of Gd-loaded liquid scintillator ($\sim$0.1\% Gd) within a transparent acrylic vessel. This will be the volume for neutrino interactions. It is surrounded by a 55 cm thick layer of unloaded scintillator (``$\gamma$-catcher'') contained in a second acrylic vessel. This scintillating volume is necessary to fully contain the energy deposition of gamma rays from the neutron capture on Gd as well as the positron annihilation gamma rays inside the central region. It also improves the rejection of the fast neutron background.
 
Surrounding the $\gamma$-catcher, a 105 cm thick region contains non-scintillating oil inside a stainless steel ``buffer'' vessel. This volume reduces by two orders of magnitude with respect to CHOOZ the level of accidental backgrounds coming mainly from the radioactivity of the photomultiplier tubes (PMTs). 390 10'' PMTs are installed on the inner wall and lids of the tank to collect the light from the central scintillating volumes, providing about 13\% photocathode coverage. 

The central detector is encapsulated within a ``inner muon veto'' shield, 50 cm thick, filled with scintillating organic liquid and instrumented with 78 8'' PMTs. It allows the identification of muons passing near the active detector that can create spallation neutrons and backgrounds coming from outside. Because of space constraint, the 70 cm sand shielding of CHOOZ is replaced by a 15 cm iron layer to protect the detector from rock radioactivity and to increase the target volume. An ``outer muon veto'' covers the top of the main system and provides additional rejection power for cosmic-induced events. It can be used for constant mutual efficiency monitoring with the inner veto.

The near and far detectors will be identical inside the PMT support structure, allowing a relative normalization error of 0.6\% or less, to be compared with the 2.7\% systematic error of the CHOOZ experiment. 

\begin{figure}[h]
\centering
\includegraphics[width=80mm]{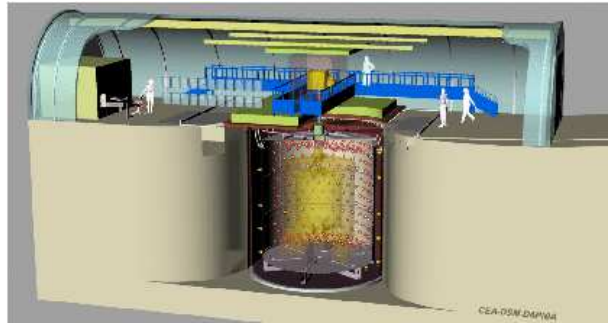}
\caption{\label{farlab}The Double Chooz Far Detector in the Experimental Hall.}
\end{figure}

Figure~\ref{farlab} shows a view of the far detector inside the experimental hall. Calibration systems will be deployed periodically into the target and $\gamma$-catcher allowing to check the stability of the system. The deployment of radioactive sources must be performed in a clean environment under a dry Nitrogen atmosphere. A glove box interface with an associated clean room will be installed on the top of the detector. The electronic racks will be located at the end of the lab.

\subsection{R\&D Activities}

An intense R\&D effort has been carried out by the Double Chooz collaboration to validate the robustness of the detector concept and the feasibility of the sensitivity goals. Some examples of the activities performed are:

\begin{itemize}
\item
The construction and operation of a 1:5 scale prototype allowed us to validate the technical choices for the vessels construction and integration of the components, check the material compatibility with the scintillator, study the liquid scintillation performances and filling system, the calibration techniques, safety and all the instrumentation needed to operate such a detector.
\item
Production and tests of stable high-quality Gd-loaded scintillator, critical for the performance of the experiment. The long-term stability, optical transmission properties, light yield, transparency and radiopurity have been tested for several years.
\item
 Magnetic measurements of the performance of the 10'' PMTs inside a controlled magnetic environment and design and optimization of individual PMT magnetic shields. 
\item
Development of 8-bit Flash ADCs waveform digitizers for the readout of each PMT signal, recording the pulse shape for possible offline particle identification studies.
\end{itemize}

\subsection{Systematic Errors and Backgrounds}

Many systematic uncertainties that affected CHOOZ and all previous single-baseline reactor neutrino experiments are greatly reduced by having both near and far detectors. Table \ref{sys} summarizes the systematic uncertainties in the measurement of the antineutrino flux comparing both CHOOZ and Double Chooz detectors.

\begin{table}[h]
\begin{center}
\caption{Systematic Uncertainties in CHOOZ and Double Chooz Reactor Experiments.} 
\begin{tabular}{|l|c|c|}
\hline                       
  & \textbf{CHOOZ} & \textbf{Double Chooz}\\
\hline
Reactor fuel cross section & 1.9\% & -- \\
Reactor power & 0.7\% & -- \\
Energy per fission & 0.6\% & -- \\
Number of protons & 0.8\% & 0.2\% \\
Detection efficiency & 1.5\% & 0.5\% \\
\hline
\textbf{TOTAL} & 2.7\% & 0.6\% \\
\hline
\end{tabular}
\label{sys}
\end{center}
\end{table}

Though an uncertainty from the neutrino contribution of spent fuel pools remains, it is negligible for Double Chooz. The neutrino rates are proportional to the number of free protons inside the target volumes, which thus has to be experimentally determined with a precision of 0.2\%. This constitutes one of the major improvements with respect to CHOOZ. In addition, a comprehensive calibration system consisting of radioactive sources deployed in different detector regions, laser light flashes and LED pulses, will be enforced to correct the unavoidable differences between the two detector responses. The optimization of the Double Chooz detector design allows us to simplify the analysis and to reduce the detection efficiency systematic errors down to 0.5\% while keeping high statistics.

The selection of high pure materials for detector construction and passive shielding around the active region provide an efficient protection against accidental background events. Furthermore, they can be measured {\it in situ} even with reactors on. The inner and outer veto systems and the inner detector muon electronics are designed to address the correlated background. The error due to the background subtraction in both detectors is expected to be less than 1\%.

\subsection{Expected Sensitivity}

\begin{figure*}[t]
\centering
\includegraphics[width=135mm]{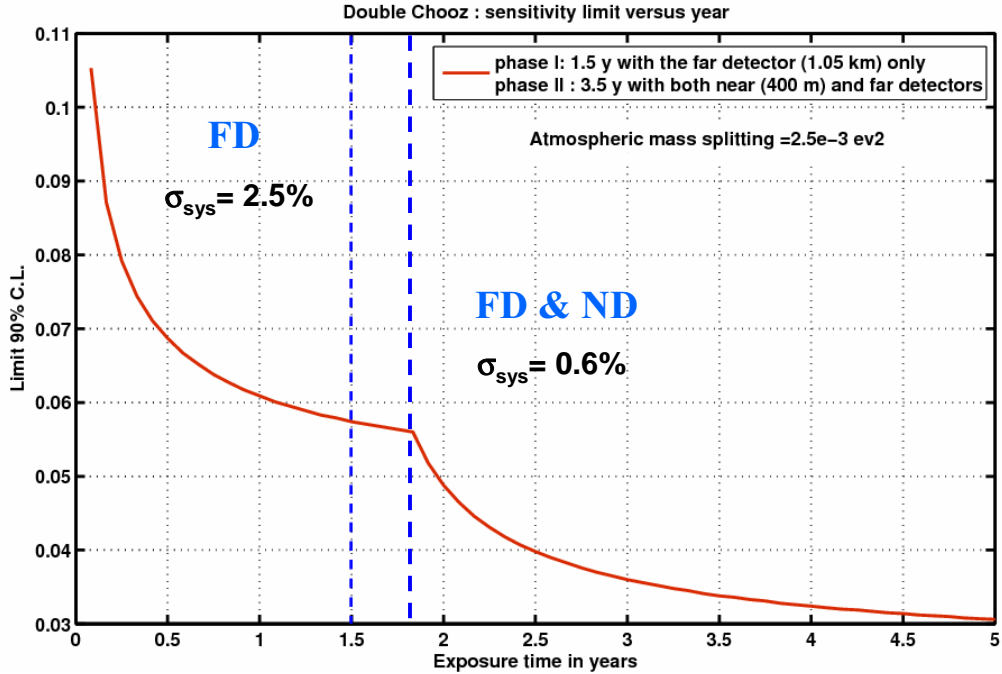}
\caption{\label{sensit}Double Chooz expected sensitivity limit (90\% C.L.) to $sin^2(2\theta_{13})$ as a function of time, assuming that the near detector is built 1.5 years after the far detector.}
\end{figure*}

The Double Chooz experiment will make a fundamental contribution to the determination of the $\theta_{13}$ mixing angle within an unrivaled time scale. In a first phase, the far detector will start taking data alone. In a few months the previous CHOOZ limit will be surpassed. After $\sim$1.5 years of data taking with one detector, Double Chooz will be sensitive to $sin^2(2\theta_{13})$ $>$ 0.06 (see Fig. \ref{sensit}), a factor 3 better than CHOOZ. 

In a second phase, with both near- and far- detectors running simultaneously, the systematic errors can be reduced down to 0.6\%. Double Chooz will explore $sin^2(2\theta_{13})$ $>$ 0.03 after three years of operation, improving the current limit by almost one order of magnitude.

\subsection{Present Status}

The Double Chooz experiment is now in the installation phase of the far detector in the existing underground laboratory at Chooz. It started in May 2008 with the integration of the external shield made of steel bars previously demagnetized at Chooz to avoid extra magnetic field contribution inside the detector. 

The scintillator production is on going and all the elements have been delivered. The mixing process will be done in a single batch for the two detectors to ensure identical proton per volume concentrations thereby avoiding different aging effects. 

\begin{figure}[h]
\centering
\includegraphics[width=80mm]{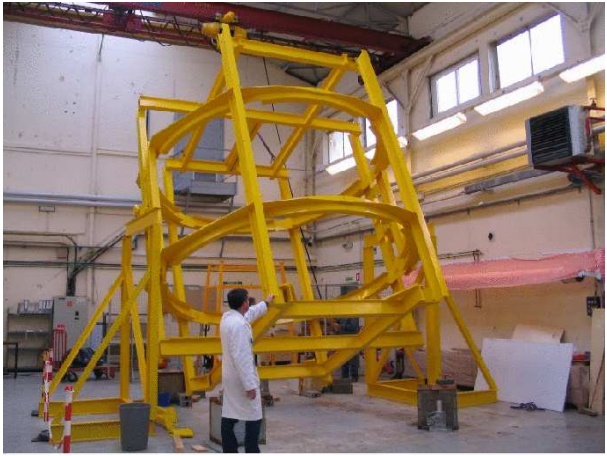}
\caption{\label{tools}Mechanical tool for the construction, transportation and installation of the $\gamma$-catcher vessel in the Chooz pit.}
\end{figure}

The design of the mechanical vessels has been completed and approved and the manufacturing is on going. The target and $\gamma$-catcher vessels will need special tools already available to be transported and installed in the pit (see Fig.~\ref{tools}).

\begin{figure}[h]
\centering
\begin{tabular}{cc}
\includegraphics[width=45mm]{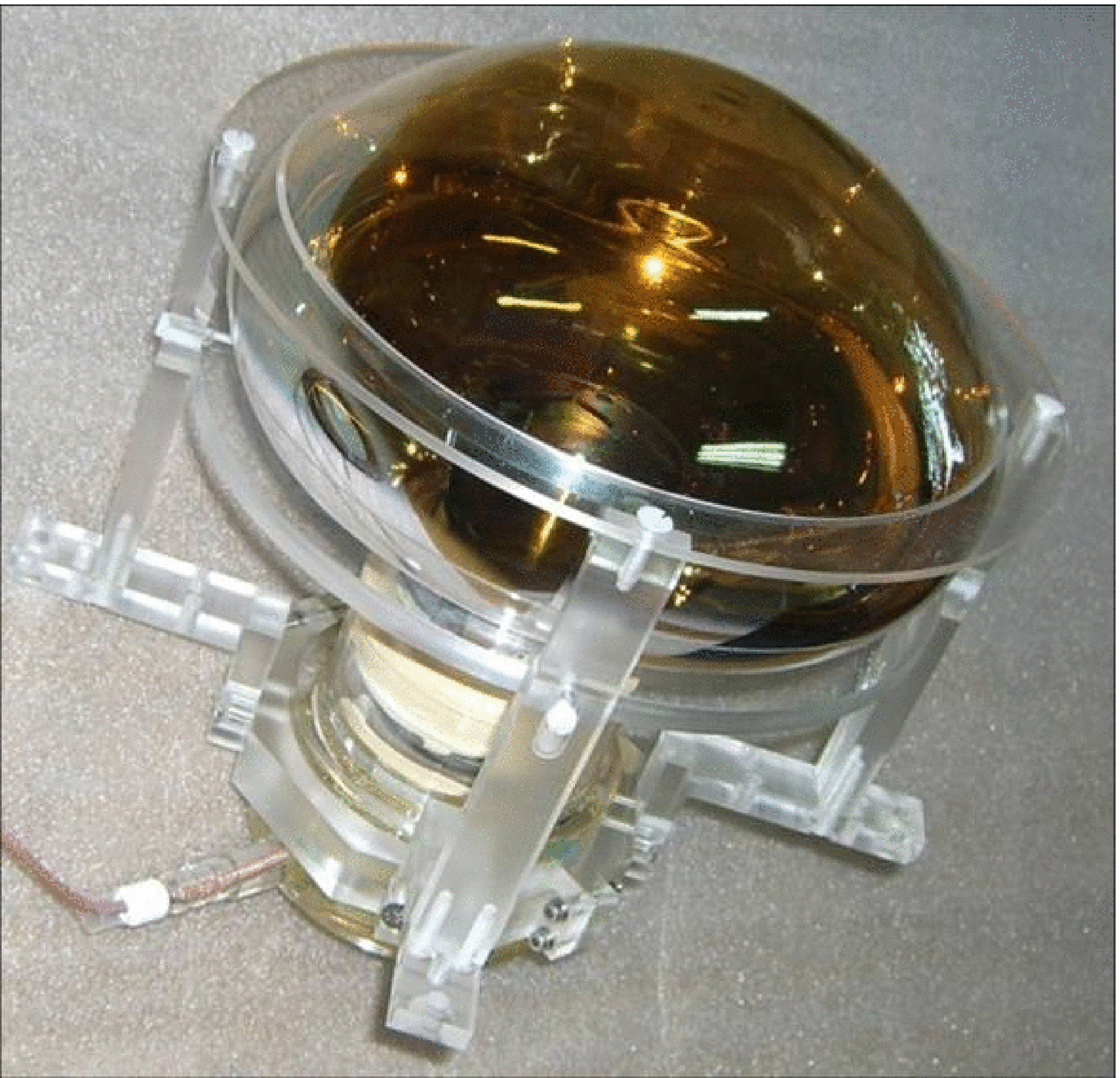} & 
\includegraphics[width=36mm]{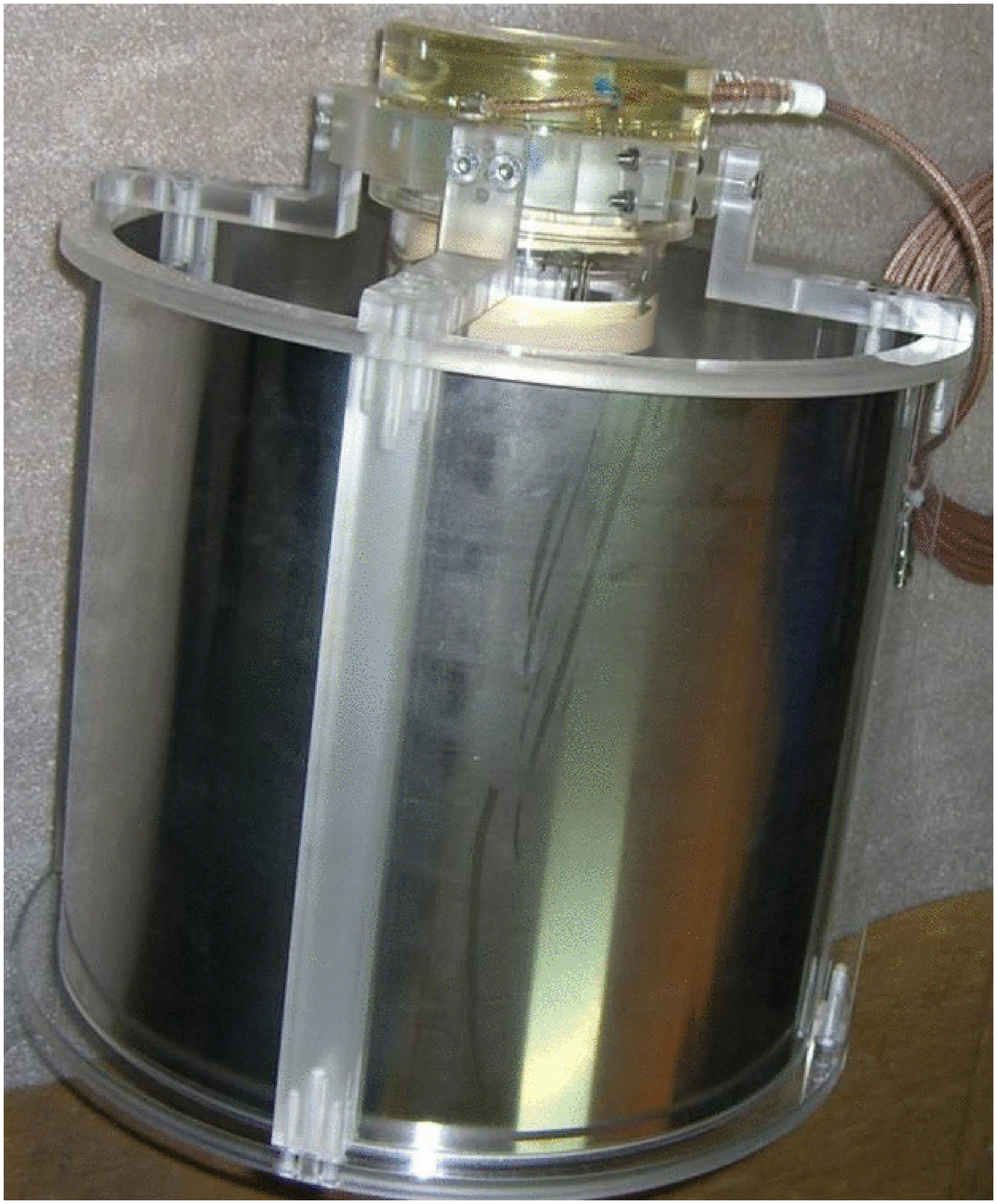}
\end{tabular}
\caption{\label{pmt}Final PMT mechanical support with the 10'' PMT and magnetic shield.}
\end{figure}

The production of the PMTs for both detectors is almost finished and a complete characterization of the buffer and inner veto PMTs is being performed in several institutions of the collaboration. The individual magnetic shields have been manufactured and their magnetic properties have been tested. The production of the acrylic mechanical supports for the buffer PMTs has finished and the assembly of all the components of the PMT system will be completed before the end of 2008. Figure~\ref{pmt} shows a picture of the final PMT system already assembled with the PMT, the mechanical support and the mu-metal shield.

The veto systems are also in production. The encapsulation process of the inner veto PMTs is on going. A prototype of the outer veto system is in construction.

A complete calibration program has been designed including the deployment of gamma and neutron radioactive sources in the different liquids of the detector with different systems depending on the volume to be accessed (guide tubes, articulated arm, fish-line, etc.) and light flashers (LEDs and lasers) to measure the PMTs. All these systems are in construction.

The far detector integration will be finished by summer 2009. The near lab site will be available at the beginning of 2010 to accommodate the second detector. In 2011, both near and far detectors, will be operative and taking data for three more years in order to achieve Double Chooz expectations.

\section{Conclusions}

Double Chooz is the first reactor neutrino experiment of a new generation using two identical detectors at different distances to measure the still unknown $\theta_{13}$ mixing angle. Double Chooz has already started its installation and it is expected to start taking data with the far detector in 2009 and with both near and far detectors in 2011.

Double Chooz will be able to measure $sin^2(2\theta_{13})$ to 3$\sigma$ if $sin^2(2\theta_{13})$ $>$ 0.05. Otherwise, it will exclude the mixing angle down to $sin^2(2\theta_{13})$ $>$ 0.03 at 90\% C.L. after three years of operation with both detectors in case of no oscillation were observed. This will represent an improvement of about a factor 7 compared to the CHOOZ limit and will open the way for a new level of accuracy in reactor neutrino experiments. The information gained with Double Chooz will complement future results with accelerator experiments, affected by degeneracy problems, helping to better constrain the last undetermined mixing parameters.

\bigskip 

\end{document}